\documentclass[11pt,numbers,sort&compress]{article}

\usepackage{latexsym,amsmath,amssymb,theorem,epsfig,graphics}
\usepackage[usenames]{color} 
\usepackage{graphicx,color}
\usepackage{fancyvrb, pstricks}

\DeclareFontFamily{OT1}{rsfs10}{}
\DeclareFontShape{OT1}{rsfs10}{m}{n}{ <-> rsfs10 }{}
\DeclareMathAlphabet{\mathscript}{OT1}{rsfs10}{m}{n}

\newcommand{\be}{\begin{equation}}
\newcommand{\ee}{\end{equation}}

\newcommand{\bea}{\begin{eqnarray}}
\newcommand{\eea}{\end{eqnarray}}
\newcommand{\ba}{\begin{array}}
\newcommand{\ea}{\end{array}}

\newcommand{\comment}[1]{}

\usepackage[all]{xy}

\begin{document}

\begin{titlepage}
  \begin{flushright}
    arXiv:0901.1662
  \end{flushright}
  \vspace*{\stretch{1}}
  \begin{center}
    \huge {\bf A Simple Introduction to Gr\"obner Basis \\Methods in
      String Phenomenology}
  \end{center}
\vspace{0.1cm}
  \begin{center}
    \begin{minipage}{\textwidth}
      \begin{center}
        {\Large  James Gray}
       \end{center}
    \end{minipage}
  \end{center}
  \vspace*{1mm}
  \begin{center}
    \begin{minipage}{\textwidth}
{\small
      \begin{center}
        \large
        Rudolf Peierls Centre for Theoretical Physics, \\
        University of Oxford.\\[0.2cm]
      \end{center}
}
    \end{minipage}
  \end{center}
  \vspace*{\stretch{1}}
  \begin{abstract}
    \normalsize 
    In this talk I give an elementary introduction to the key
    algorithm used in recent applications of computational algebraic
    geometry to the subject of string phenomenology. I begin with a
    simple description of the algorithm itself and then give 3
    examples of its use in physics. I describe how it can be used to
    obtain constraints on flux parameters, how it can simplify the
    equations describing vacua in 4d string models and lastly how it
    can be used to compute the vacuum space of the electroweak sector
    of the MSSM.
  \end{abstract}
  \vspace*{\stretch{5}}
  \begin{minipage}{\textwidth}
    \underline{\hspace{5cm}}
    \\
    \footnotesize email: j.gray1@physics.ox.ac.uk
  \end{minipage}
\end{titlepage}


\section{Introduction}

There is currently a great deal of interest in applying the methods of
computational algebraic geometry to string phenomenology and closely
related sub-fields of theoretical physics. For some examples of recent
work see
\cite{VacspaceBlock,Benvenuti:2006qr,Feng:2007ur,Forcella:2008bb,GIOBlock1,Ferrari:2008rz,Distler:2005hi,StringvacuaBlock,Font:2008vd,ModelBuildingBlock,Kaura:2006mv,Raby:2007yc,NumericMetricBlock,Candelas:2008wb}
and references therein. These papers utilise advances in algorithmic
techniques in commutative algebra to study a wide range of subjects
including various aspects of globally supersymmetric gauge theory
\cite{VacspaceBlock,Benvenuti:2006qr,Feng:2007ur,Forcella:2008bb,GIOBlock1,Ferrari:2008rz},
finding flux vacua in string phenomenology
\cite{Distler:2005hi,StringvacuaBlock,Font:2008vd}, studying heterotic
model building on smooth Calabi-Yau in non-standard embeddings
\cite{ModelBuildingBlock} and more besides
\cite{Kaura:2006mv,Raby:2007yc,NumericMetricBlock,Candelas:2008wb}.

Despite the wide range of physical problems which have been addressed
within this context, the computational tools which are being used are
all based, finally, on the same algorithm. The Buchberger algorithm
\cite{Buchberger} is at once what lends these methods their power and
also the rate limiting step - placing bounds on the size of problem
that can be dealt with. The recent burst of activity in this field has
been fueled, in part, by the advent of freely available, efficient
implementations of this algorithm \cite{m2,sing}. There are also
interfaces available between the commutative algebra program
\cite{sing} and Mathematica \cite{singm,StringvacuaBlock}, with
\cite{StringvacuaBlock} being particularly geared towards physicist's
needs. The aim of this talk is to give an elementary introduction to
the Buchberger algorithm and some of its recent applications.

In order to give an idea of how one simple algorithm can make so much
possible, I will present the Buchberger algorithm and then show how it
may be applied to physics in 3 elementary examples. Firstly, I will
describe how it can be used to obtain constraints on the flux
parameters in four dimensional descriptions of string phenomenological
models which are necessary and sufficient for the existence of certain
types of vacuum \cite{StringvacuaBlock}. Secondly, I will describe how
the Buchberger algorithm can be used to simplify the equations
describing the vacua of such systems - making problems of finding
minima much more tractable \cite{StringvacuaBlock}. Finally, I shall
describe how the same simple algorithm can be used to calculate the
supersymmetric vacuum space geometry of the electroweak sector of the
MSSM \cite{VacspaceBlock}.

The remainder of this talk is structured as follows. In the next
section, I take a few pages to explain the algorithm and the few
mathematical concepts that we will require. In the three sections
following that, I then describe the three examples mentioned above. I
shall conclude by making a few final comments about the versatility
and scaling of the Buchberger algorithm.

\section{A tiny bit of commutative algebra}
\label{math}

Two pages of simple mathematics will suffice to achieve all of the
physical goals mentioned in the introduction. First of all we define
the notion of a polynomial ring. In this paper we will call the fields
of the physical systems we study $\phi^i$ and any parameters present,
such as flux parameters, $a^{\alpha}$. The polynomial rings
$\mathbb{C}\left[ \phi^i ,a^{\alpha} \right]$ and
$\mathbb{C}\left[a^{\alpha} \right]$ are then simply the infinite set
of all polynomials in the fields and parameters, and the infinite set
of all polynomials in the parameters respectively.

Another mathematical concept we will require is that of a monomial
ordering. This is simply an unambiguous way of stating whether any
given monomial is formally bigger than any other given monomial. We
may denote this in a particular case by saying $m_1 > m_2$ where $m_1,
m_2 \in \mathbb{C} \left[ \phi^i, a^{\alpha} \right]$ are monomials in
the fields and parameters. It is important to say what is {\it not}
meant by this. We are not saying that we are taking values of the
variables such that the monomial $m_1$ is numerically larger than the
monomial $m_2$. Rather we are saying that, in our formal ordering,
$m_1$ is considered to come before $m_2$.

For our purposes we will require a special type of monomial ordering
called an elimination ordering. This means that our formal ordering of
monomials has the following property.  \bea P \in \mathbb{C} \left[
  \phi^i, a^{\alpha} \right] , \textnormal{LM}(P) \in
\mathbb{C}\left[a^{\alpha}\right] \Rightarrow P \in \mathbb{C}\left[
  a^{\alpha} \right] \eea In words this just says that if the largest
monomial in $P$ according to our ordering, $\textnormal{LM}(P)$, does
not depend on $\phi^i$ then $P$ does not depend on the fields at
all. The monomial ordering classes all monomials with fields in them
as being bigger than all of those without such constituents.

Given this notion of monomial orderings we can now present the one
algorithm we will need to use - the Buchberger algorithm
\cite{Buchberger}. The Buchberger algorithm takes as its input a set
of polynomials. These may be thought of as a system of polynomial
equations by the simple expedient of setting all of the polynomials to
zero. The algorithm returns a new set of polynomials which, when
thought of as a system of equations in the same way, has the same
solution set as the input. The output system, however, has several
additional useful properties as we will see.

\subsection*{The Buchberger Algorithm}
\begin{enumerate}
\item Start with a set of polynomials, call this set ${\cal G}$.
\item Choose a monomial ordering with the elimination property
  described above.
\item For any pair of polynomials $P_i, P_j \in {\cal G}$ multiply by
  monomials and form a difference so as to cancel the leading
  monomials with respect to the monomial ordering: \bea S = p_1 P_I -
  p_2 P_J \;\; \textnormal{s.t.} \;\; p_1 \textnormal{LM}(P_I), p_2
  \textnormal{LM}(P_J) \;\; \textnormal{cancel} \;.\eea
\item Perform polynomial long division of $S$ with respect to ${\cal
    G}$. That is, form $\tilde{h} = S - m_3 P_k$ where $m_3$ is a
  monomial and $P_k \in {\cal G}$ such that $m_3 \textnormal{LM}(P_k)$
  cancels a monomial in $S$. Repeat until no further reduction is
  possible. Call the result $h$.
\item If $h=0$ consider the next pair. If $h \neq 0$ add $h$ to ${\cal
    G}$ and return to step 3.
\end{enumerate}
The algorithm terminates when all S-polynomials which may be formed
reduce to $0$. The final set of polynomials is called a Gr\"obner
basis.

As mentioned above, the resulting set of polynomials has several nice
properties. The feature which is often taken as defining is that
polynomial long division with respect to this new set of polynomials
always gives the same answer - it does not matter in which order we
divide the polynomials out by.

For us, however, the important point about our Gr\"obner basis ${\cal
  G}$ is that it has what is called the elimination property. ${\cal
  G} \cap \mathbb{C} \left[ a^{\alpha} \right]$, the set of all
polynomials in ${\cal G}$ which depend only upon the parameters, gives
a complete set of equations on the $a^{\alpha}$ which are necessary
and sufficient for the existence of a solution to the set of equations
we started with. The reason why this is so is actually very
straightforward. Our elimination ordering says that any monomial with
a field in it is greater than any monomial only made up of
parameters. Looking back at step 3 of the Buchberger algorithm we see
that we are repeatedly canceling off the leading terms of our
polynomials - those containing the fields - as much as we can. Thus if
it is possible to rearrange our initial equations to get expressions
which do not depend upon the fields $\phi^i$ then the Buchberger
algorithm will do this for us. Clearly, while we have interpreted the
$a^{\alpha}$ as parameters and the $\phi^i$ as fields in the above, as
this is what we will require for the next section, this was not
necessary. The Buchberger algorithm can be used to eliminate any
unwanted set of variables from a problem, in the manner we have
described.

This completes all of the mathematics we will need for our entire
discussion and we may now move on to apply what we have learnt.

\section{Constraints}

The first physical question we wish to answer is the following. Given
a four dimensional ${\cal N}=1$ supergravity describing a flux
compactification, what are the constraints on the flux parameters
which are necessary and sufficient for the existence of a particular
kind of vacuum? This question can be asked, and answered
\cite{StringvacuaBlock}, for any kind of vacuum, but in the interests
of concreteness and brevity let us restrict ourselves to the simple
case of supersymmetric Minkowski vacua.

Here is the superpotential of a typical system, taken from
\cite{Shelton:2005cf}. It describes a non-geometric compactification
of type IIB string theory.
\begin{eqnarray} \label{sheltonW}
W & = & a_0 - 3a_1 \tau + 3a_2 \tau^2 - a_3 \tau^3\\ \nonumber
& &
 \hspace{0.2in} + S (-b_0 + 3b_1 \tau - 3b_2 \tau^2 + b_3 \tau^3)
\nonumber\\ & &
 \hspace{0.2in} + 3 U (c_0 + (\hat{c}_1 +\check{ c}_1 + \tilde{c}_1)
 \tau - (\hat{c}_2 +\check{ c}_2 + \tilde{c}_2) \tau^2 -c_3
 \tau^3), \nonumber
\end{eqnarray}
This system has some known constraints on its parameters which are
necessary for the existence of a permissible vacuum. These come from,
for example, tadpole cancellation conditions.  \bea\label{constSTW}
a_0 b_3-3 a_1 b_2+3 a_2b_1-a_3b_0 = 16 \, \, \,\, && \\ \nonumber a_0
c_3+a_1 (\check{c}_2 + \hat{c}_2-\tilde{c}_2) -a_2 (\check{c}_1 +
\hat{c}_1 -\tilde{c}_1)-a_3c_0 = 0 \,\,\,\, && \\ \nonumber \ba{ccc}
\ba{rcl}
c_0 b_2-\tilde{c}_1 b_1+\hat{c}_1 b_1-\check{c}_2 b_0 & = & 0\\
\check{c}_1 b_3-\hat{c}_2 b_2+\tilde{c}_2 b_2-c_3 b_1 & = & 0\\
c_0 b_3-\tilde{c}_1 b_2+ \hat{c}_1 b_2-\check{c}_2 b_1 & = & 0\\
\check{c}_1 b_2-\hat{c}_2 b_1+\tilde{c}_2 b_1-c_3 b_0 & = & 0\\
\ea & \ba{rcl}
c_0\tilde{c}_2-\check{c}_1 ^ 2+\tilde{c}_1\hat{c}_1-\hat{c}_2 c_0 & = & 0\\
c_3\tilde{c}_1-\check{c}_2 ^ 2 +\tilde{c}_2\hat{c}_2-\hat{c}_1 c_3 & = & 0\\
c_3 c_0-\check{c}_2\hat{c}_1
+\tilde{c}_2\check{c}_1-\hat{c}_1\tilde{c}_2 & = & 0\\
\hat{c}_2\tilde{c}_1-\tilde{c}_1\check{c}_2
+\check{c}_1\hat{c}_2-c_0c_3 & =& 0 \ . \\
\ea \ea \eea We also have the same constraints with the hats and
checks switched around. In this example the fields, which we have been
calling $\phi^i$, are $S,\tau$ and $U$ and everything else is a
``flux'' parameter, or an $a^{\alpha}$ in our notation.

In total, the equations which must be satisfied if a supersymmetric
Minkowski vacuum is to exist are $W=0$, $\partial_S W=0$,
$\partial_{\tau} W=0$, $\partial_U W=0$ and the constraints on the
flux parameters given above. To extract a set of constraints solely
involving the parameters which are necessary and sufficient for the
existence of a solution to these equations we simply follow the
procedure outlined in the previous section.

We can carry out this calculation trivially in Stringvacua
\cite{StringvacuaBlock} and, in fact, this example is provided for the
user in the help system.  The result is

\bea 0 &=& \tilde{c}_1 = \tilde{c}_2 = \hat{c}_1 = \hat{c}_2 =
\check{c}_1 = \check{c}_2 = c_0 = c_3 \\ \nonumber 0 &=& 16+a_3 b_0-3
a_2 b_1+3 a_1 b_2-a_0 b_3 \\ \nonumber 0 &=& 16 a_3^2 b_0^2-96 a_2 a_3
b_0 b_1-288 a_2^2 b_1^2+432 a_1 a_3 b_1^2+54 a_2^3 b_1^3-81 a_1 a_2
a_3 b_1^3+27 a_0 a_3^2 b_1^3 +432 a_1 a_3 b_0 b_2 \\ \nonumber && -27
a_2^2 a_3 b_0^2 b_2+48 a_1 a_3^2 b_0^2 b_2-288 a_0 a_3 b_1 b_2-18 a_1
a_2 a_3 b_0 b_1 b_2-45 a_0 a_3^2 b_0 b_1 b_2 -54 a_1 a_2^2 b_1^2 b_2\\
\nonumber && +81 a_1^2 a_3 b_1^2 b_2-27 a_0 a_2 a_3 b_1^2 b_2+54 a_0
a_2 a_3 b_0 b_2^2+27 a_0 a_1 a_3 b_1 \ b_2^2-27 a_0^2 a_3 b_2^3-288
a_1 a_2 b_0 b_3 \\ \nonumber && -32 a_0 a_3 b_0 b_3+27 a_2^3 b_0^2 \
b_3-45 a_1 a_2 a_3 b_0^2 b_3+432 a_0 a_2 b_1 b_3-27 a_1 a_2^2 b_0 b_1
b_3+54 a_1^2 a_3 b_0 b_1 b_3 \\ \nonumber && +48 a_0 a_2 a_3 b_0 b_1
b_3 +18 a_0 a_2^2 b_1^2 b_3-81 a_0 a_1 a_3 b_1^2 b_3-144 a_0 a_1 b_2
b_3+27 a_1^2 a_2
b_0 b_2 b_3-54 a_0 a_2^2 b_0 b_2 b_3 \\
\nonumber && -51 a_0 a_1 a_3 b_0 b_2 b_3 +27 a_0 a_1 a_2 b_1 b_2
b_3+45 a_0^2 a_3 b_1 b_2 b_3-27 a_0 a_1^2 b_2^2 b_3+27 a_0^2 a_2 b_2^2
b_3+16 a_0^2 b_3^2 \\
\nonumber && -27 a_1^3 b_0 b_3^2+45 a_0 a_1 a_2 b_0 b_3^2 +27 a_0
a_1^2 b_1 b_3^2-48 a_0^2 a_2 b_1 b_3^2+3 a_0^2 a_1 b_2 b_3^2 \eea

The reader will note that the result is a somewhat lengthy set of
equations. In principle one has to find quantized solutions to these
expressions, an obviously intractable Diophantine problem, and
therefore it might be asked why this result is of any use. In fact,
knowledge of such constraints on the flux parameters is hugely useful
for a number of reasons.

\begin{itemize}
\item Firstly, we note that, while the full result of this process is
  often complex, some of the constraints can give us simple
  information about the system. In the current case, for example it
  can be seen that $\tilde{c}_2=0$ is required for the existence of a
  supersymmetric Minkowski vacuum.
\item Secondly, if one is scanning over a range of flux parameters and
  trying to numerically solve the equations to find vacua one can
  speed up ones analysis by first substituting any given set of flux
  parameters into the constraints we have obtained. If the constraints
  are not satisfied then vacua do not exist and there is no point in
  searching numerically for a solution. This turns what would be a
  time consuming numerical process giving inconclusive results (no
  solution was found) into a quick analytic conclusion (no solution
  exists).
\item Lastly, knowledge of such constraints can greatly speed up
  algebraic approaches to finding vacua such as those outlined in
  \cite{StringvacuaBlock}.
\end{itemize}

\section{Simplifying equations for vacua}

Another use for the mathematics we learnt in section \ref{math} are
the so called ``splitting tools'' used in work such as
\cite{StringvacuaBlock}.  The physical idea here is simple. Consider
trying to solve the equations $\partial V / \partial \phi^i =0$ to
find the vacua, including those which spontaneously break
supersymmetry, of some supergravity theory. These equations are often
extremely complicated. One way of viewing why this is so is that the
equations for the turning points of the potential contain a {\it lot}
of information. They describe not only the isolated minima of the
potential which are of interest, but also lines of maxima, saddle
points of various sorts and so forth. A useful tool to have,
therefore, would be an algorithm that takes such a system as an input
and returns a whole series of separate sets of equations, each
individually describing fewer turning points. Since each separate
equation system would then contain less information one might expect
them to be easier to solve. It would be beneficial to choose a
division of these equations which has physical interest. The choice we
will make here, and which programs like Stringvacua implement
\cite{StringvacuaBlock}, is to split up the equations for the turning
points according to how they break supersymmetry - that is according
to which F-terms vanish when evaluated on those loci.

The ability that packages such as Stringvacua have to split up
equations in this manner is based upon the following splitting tool
(see \cite{Stillman} for a nice set of more detailed notes on this
kind of mathematical technique). Say that one of the F-terms of our
theory is called $F$. Then we can split the equations describing
turning points of the potential into two pieces.
\bea \label{splitone}
\partial V / \partial \phi^i =0 \;,\; F = 0 \\ \label{neqpart}
\partial V/ \partial \phi^i = 0 \;,\; F \neq 0
\eea

The first of these expressions is a set of equations which is easier
to solve, in general, than $\partial V / \partial \phi^i =0$ alone. We
can use the F-term to simplify the equations for the turning points of
the potential. On the other hand, expression \eqref{neqpart} is not
even a set of equations - it contains an inequality. We can convert
\eqref{neqpart} into a system purely involving equalities by making
use of the mathematics we learned in section \ref{math}.

Consider the following set of equations, including a dummy variable $t$.
\bea \label{splitconsider}
\partial V/\partial \phi^i =0 \; ,\; F t-1 =0 \eea The second equation
in \eqref{splitconsider} has a solution if and only if $F \neq 0$,
simply $t = 1/F$. If $F=0$ the equation reduces to $-1=0$ which
clearly has no solutions.  The equations \eqref{splitconsider}, then,
have a solution whenever the set of equalities and inequalities
\eqref{neqpart} do. Unfortunately they also depend upon one extra, and
unwanted, variable - $t$. This is not a problem as we already know how
to remove unwanted variables from our equations. We can simply
eliminate them, as we did the fields in section \ref{math}. This will
leave us with a necessary and sufficient set of equations in $\phi^i$
and $a^{\alpha}$ for a solution to \eqref{splitconsider} and thus to
\eqref{neqpart}.

So we can split the equations for the turning points of our potential
into two simpler systems. One describes the turning points of $V$ for
which $F=0$ and the other those for which $F \neq 0$. We can of course
perform such a splitting many times - once for each F-term! In fact,
using additional techniques from algorithmic algebraic geometry
\cite{primdec,StringvacuaBlock}, which are essentially based upon the
same trick, one can go much further. One can split the equations for
the turning points up into component parts gaining one set of
equations for every separate locus. Because we know which F-terms are
non-zero on each of them these are classified according to how they
break supersymmetry. The researcher interested in a certain type of
breaking can therefore select the equations describing the vacua of
interest and throw everything else away.

The above process of splitting up the equations for the vacua of a
system can be very simply carried out in Stringvacua. Numerous
examples can be found in the Mathematica help files which come with
the package \cite{StringvacuaBlock}. Here, let us consider the example
of M-theory compactified on the coset $\frac{SU(3) \times U(1)}{U(1)
  \times U(1)}$. The K\"ahler and superpotential for this coset, which
has $SU(3)$ structure, has been presented in \cite{Micu:2006ey}.
\begin{eqnarray}
  K &=& -4 \log (-i(U- \bar{U})) - \log (-i (T_1-\bar{T}_1) (T_2
  -\bar{T}_2) (T_3 - \bar{T}_3))  \\ \nonumber
  W &=& \frac{1}{\sqrt{8}} \left[ 4 U (T_1+T_2+T_3) + 2 T_2 T_3 - T_1 T_3
    - T_1 T_2 + 200  \right] 
\end{eqnarray}
Even this, relatively simple, model results in a potential of
considerable size. Defining $T_i = -i t_i + \tau_i$ and $U = -i x + y$
we find,
\begin{eqnarray} \label{bigchap}
  \nonumber
  V&=& \frac{1}{256 t_1 t_2 t_3 x^4} (40000 + t_3^2 \tau_1^2 -
  400 \tau_1 \tau_2 - 4 t_3^2 \tau_1 \tau_2 + 4 t_3^2 \tau_2^2 +
  \tau_1^2 \tau_2^2 - 400 \tau_1 \tau_3 + 800 \tau_2 \tau_3 +
  \\ && 2 \tau_1^2 \tau_2 \tau_3 - 4 \tau_1 \tau_2^2
  \tau_3  + \tau_1^2 \tau_3^2   
  - 4 \tau_1 \tau_2 \tau_3^2 + 4 \tau_2^2 \tau_3^2 - 24 t_2 t_3 x^2 +
  4 t_3^2 x^2 - 24 t_1 (t_2 + t_3) x^2 \\ \nonumber && + 4 \tau_1^2 x^2 +
  8 \tau_1 \tau_2 x^2 + 4 \tau_2^2 x^2 + 8  \tau_1  \tau_3  x^2   
  + 8  \tau_2  \tau_3  x^2 + 4  \tau_3^2  x^2 + 1600  \tau_1  y -
  8  t_3^2  \tau_1  y \\ \nonumber && + 1600  \tau_2  y + 16
  t_3^2  \tau_2  y - 
  8  \tau_1^2  \tau_2  y - 8  \tau_1  \tau_2^2  y + 1600 \tau_3  y -
  8  \tau_1^2  \tau_3  y + 16  \tau_2^2  \tau_3  y - 8 \tau_1
  \tau_3^2  y  
  \\ \nonumber && + 16  \tau_2  \tau_3^2  y + 16  t_3^2  y^2 + 16
  \tau_1^2  y^2 + 
  32  \tau_1  \tau_2  y^2   + 16  \tau_2^2  y^2 + 32  \tau_1
  \tau_3  y^2 + 
  32  \tau_2  \tau_3  y^2 + 16  \tau_3^2  y^2 \\ \nonumber && 
  + t_1^2   (t_2^2 + t_3^2 + \tau_2^2 + 2  \tau_2  \tau_3 + \tau_3^2 + 
  4  x^2  - 8 \tau_2  y - 8  \tau_3  y + 16  y^2 ) +
  t_2^2   (4  t_3^2 + \tau_1^2 - 4  \tau_1  (\tau_3 + 2  y ) \\
  \nonumber && + 
  4   (\tau_3^2 + x^2 + 4 \tau_3  y + 4  y^2)) \;.
\end{eqnarray}

To find the turning points of this potential we naively need to take
eight different derivatives of \eqref{bigchap} and solve the resulting
set of inter-coupled equations in eight variables. This is clearly
prohibitively difficult. Using the techniques described in this
section, however, Stringvacua, can separate off parts of the vacuum
space for us with ease. Consider, for example, the vacua which are
isolated in field space and for which the real parts of all of the
F-terms are non-zero, with the imaginary parts vanishing. To find
these, the package tells us, we need only solve the equations, \bea 9
x^2 - 500 = 0 \;,\; 5 t_1-2 x =0 \;,\; t_2-x=0 \;,\; t_3-x=0 \;,\;
\tau_1=\tau_2=\tau_3=y=0 \;. \eea Because they only describe a small
subset of all of the turning points of the full potential these
equations are extremely simple in form and may be trivially
solved. For this particular example the physically acceptable turning
point that results is a saddle - something which can be readily
ascertained once its location has been discovered.

\section{Geometry of vacuum spaces}

As a final example of what we can do with the simple techniques
introduced in section \ref{math} we will show how to calculate the
vacuum space of a globally supersymmetric gauge theory. It is a well
known result (see \cite{Luty:1995sd} and references therein) that the
supersymmetric vacuum space of a such a theory, with gauge group $G$,
can be described as the space of holomorphic gauge invariant operators
(GIO's) built out of F-flat field configurations. What does this space
look like?  Consider a space, the coordinates of which are identified
with the GIO's of the theory. If there were no relations amongst the
gauge invariant operators then this space would be the vacuum
space. However there frequently are relations because of the way in
which the GIO's are built out of the fields. For example, if we have
three gauge invariant operators $S^1,S^2$ and $S^3$ which are built
out of the fields as $S^1 = (\phi^1)^2, S^2 = (\phi^2)^2 , S^3 =
\phi^1 \phi^2$ then we have the relation $S^1 S^2 = (S^3)^2$. If we
take these GIO's to be built out of the F-flat field configurations
then there will be still further relations among them. The vacuum
space of the theory is the subspace defined by the solutions of these
equations describing relations amongst the gauge invariant operators,
once F-flatness has been taken into account.

How can we calculate such a thing? The holomorphic gauge invariant
operators of a globally supersymmetric gauge theory are given in terms
of the fields.  \bea S^I = f^I (\phi^i) \eea Here $S^I$ are our GIO's
and the $f^I$ are the functions of the fields that define them. Let us
write the F-terms of the theory as $F^i$. Consider the following set
of equations.  \bea \label{finalcase} F^i =0 \;,\; S^I - f^I(\phi^i)
=0 \eea These equations have solutions whenever the $S^I$ are given by
functions of the fields in the correct way and when those field
configurations which are being used are F-flat. However, according to
the proceeding discussion, we wish to simply have equations in terms
of the GIO's to describe our vacuum space. As in previous sections, we
can eliminate the unwanted variables in our problem, in this case the
fields $\phi^i$, using the algorithm of section \ref{math} to obtain
the equations describing the vacuum space.

As a simple example, let us take the electroweak sector of the MSSM
\cite{VacspaceBlock} (with right handed neutrinos). Given the field
content of the left handed leptons, $L^i_{\alpha}$, the right handed
leptons, $e^i$ and $\nu^i$, and the two Higgs, $H$ and $\bar{H}$, one
can build the elementary GIO's given in table \ref{table1}. The
indices $i,j$ run over the 3 flavours and the indices $\alpha, \beta$
label the fundamental of $SU(2)$.

\begin{table}
{\begin{center}
\begin{tabular}{|c||c|c|c|}\hline
\mbox{Type} & \mbox{Explicit Sum} & \mbox{Index} & \mbox{Number} \\
\hline \hline
$LH$  & $L^i_\alpha H_\beta \epsilon^{\alpha \beta}$ & $i=1,2,3$ & 3 \\ \hline
$H\bar{H}$ & $H_\alpha \bar{H}_\beta \epsilon^{\alpha \beta}$ & & 1  \\ \hline
$LLe$ & $L^i_\alpha L^j_\beta e^k \epsilon^{\alpha \beta}$ & $i,j=1,2,3;
k=1,\ldots,j-1$ & 9  \\ \hline
$L\bar{H} e$ & $L^i_\alpha \bar{H}_\beta \epsilon^{\alpha \beta} e^j$ & $i,j=1,2,3$ & 9 \\
\hline
$\nu$ & $\nu^i$ & $i=1,2,3$ & 1 \\ \hline
\end{tabular}
\end{center}}{\caption{\label{table1}{\bf The set of elementary gauge
    invariant operators for the electroweak sector of the MSSM}.}}
\end{table}

To compute the F-terms we require the superpotential. Let us take the
most general renormalizable form which is compatible with the symmetries of the
theory and R-parity. 
\begin{equation}\label{renorm-ew-nu}
  W_{\rm minimal} = C^0 \sum_{\alpha, \beta} H^\alpha \bar{H}^\beta \epsilon_{\alpha \beta} + \sum_{i,j} C^3_{ij} e^i \sum_{\alpha, \beta} L^j_{\alpha} \bar{H}_\beta \epsilon^{\alpha \beta} +
  \sum_{i,j} C^4_{ij} \nu^i \nu^j + \sum_i C^5_{ij} \nu^i \sum_{\alpha, \beta} L^j_\alpha H_\beta \epsilon^{\alpha \beta}.
\end{equation}
Here $\epsilon$ is the invariant tensor of $SU(2)$ and $C^0, C^3_{ij},
C^4_{ij}, \textnormal{and}, C^5_{ij}$ are constant coefficients.

We now just follow the procedure outlined at the start of this
section. We calculate the F-terms by taking derivatives of the
superpotential, we label the gauge invariant operators $S_1$ to
$S_{23}$, we form the equations \eqref{finalcase} and then simply run
the elimination algorithm given in section \ref{math}.

The result is, upon simplification, given by six quadratic equations
in 6 variables. It is a simple description of an affine version of a
famous algebraic variety - the Veronese surface
\cite{VacspaceBlock}. What can be done with such a result? The
first observation we can make is that this vacuum space is not a
Calabi-Yau. This means, for example, that one can say definitively
that it is not possible to engineer this theory by placing a single D3
brane on a singularity in a Calabi-Yau manifold, without having to get
into any details of model building.

Secondly one can study such vacuum spaces in the hope of finding hints
at the structure of the theory's higher energy origins. In the case we
have studied in this section, for example, we can ``projectivize''
(pretend the GIO's are homogeneous coordinates on projective space
rather than flat space coordinates) and study the Hodge diamond of the
result. The structure of supersymmetric field theory tells us that
this Hodge diamond should depend on 4 arbitrary integers, but there is
nothing at low energies which prevents us from building theories with
any such integers we like. Interestingly, in the case of electroweak
theory, these integers are all zero or one.

\begin{equation}
  h^{p,q} \quad = \quad
  {\begin{array}{ccccc}
      &&h^{0,0}&& \\
      &h^{0,1}&&h^{0,1}& \\
      h^{0,2}&&h^{1,1}&&h^{0,2} \\
      &h^{0,1}&&h^{0,1}& \\
      &&h^{0,0}&& \\
\end{array}}
\quad = \quad
{\begin{array}{ccccc}
&&1&& \\
&0&&0& \\
0&&1&&0 \\
&0&&0& \\
&&1&& \\
\end{array}}.
\label{hodge}
\end{equation}

Whether this structure is indeed a hint of some high energy antecedent
or just a reflection of the simplicity of the theory is
debatable. This example does, however, demonstrate the idea of
searching for such evidence of new physics in vacuum space
structure. We should also add here that similar techniques can be used
to show that the vacuum space of SQCD is a Calabi-Yau
\cite{GIOBlock1}.

\section{Final Comments}

To conclude we shall make several points - one of which is a note of
caution, with the rest being more optimistic. The first point which we
shall make is that we should be careful lest the above discussion
makes the algorithm we have been describing sound like an all-powerful
tool. There is, as ever, a catch. In this case it is the way the
algorithm scales with the complexity of the problem. A ``worst case''
upper bound for the degree of the polynomials in a {\it reduced}
Gr\"obner basis can be found in \cite{MollerMora84}. If $d$ is the
largest degree found in your original set of equations then this bound
is, \bea \label{scaling} 2 \left( \frac{d^2}{2} + d \right)^{2^{n-1}}
\;, \eea where $n$ is the number of variables. This {\it worst case}
bound is therefore scaling doubly exponentially in the number of
degrees of freedom. These very high degree polynomials are an
indication that the problem is becoming very complex and thus
computationally intensive. Despite this, physically useful cases can
be analysed using this algorithm quickly, as demonstrated in this talk
and in the references. This scaling does mean that one is not likely
to gain much by putting one's problem on a much faster computer. One
good point about \eqref{scaling} is that if you can find a way, using
physical insight, to simplify the problem under study, then what you
can achieve may improve doubly exponentially. Such a piece of physical
insight was one of the keystones of the application of these methods
to finding flux vacua \cite{StringvacuaBlock}.

We finish by commenting that the methods of computational
commutative algebra which we have discussed here are extremely
versatile. We have been able to perform three very different tasks
simply utilizing one algorithm in a very simple manner. These methods
are of great utility in problems taken from the literature and their
implementation in a user friendly way in Stringvacua means that they
may be tried out on any given problem with very little expenditure of
time and effort by the researcher. Many more techniques from the field
of algorithmic commutative algebra could be applied to physical
systems than those described here, or indeed in the physics
literature. We can therefore expect that this subject will only
increase in importance in the future.

\section*{Acknowledgements}

The author is funded by STFC and would like to thank the University of
Pennsylvania for generous hospitality while some of this document was
being written. In addition he would like to thank the organisers of
the 2008 Vienna ESI workshop ``Mathematical Challenges in String
Phenomenology'' where the talk upon which these notes are based was
first given.

The author would like to offer heartfelt thanks to his collaborators
on the various projects upon which this talk is based. These include
Lara Anderson, Daniel Grayson, Amihay Hanany, Yang-Hui He, Anton
Ilderton, Vishnu Jejjala, Andr\'e Lukas, Noppadol Mekareeya and Brent
Nelson.


\end{document}